\documentclass[prb,twocolumn,showpacs,preprintnumbers,amsmath,amssymb,superscriptaddress]{revtex4}
\usepackage{graphicx}% Include figure files
\usepackage{dcolumn}% Align table columns on decimal point
\usepackage{bm}% bold math

\begin{document}

\title{Tunable terahertz oscillations in superlattices subject to in-plane magnetic field}
\author{M. Orlita}
 \email{orlita@karlov.mff.cuni.cz}
\affiliation{Charles University, Faculty of Mathematics and Physics, Institute
of Physics, Ke Karlovu 5,
 CZ-121~16 Prague 2, Czech Republic}
\affiliation{Institute of Physics, Academy of Sciences of the Czech Republic,
Cukrovarnick\'{a} 10, CZ-162 53 Prague 6, Czech Republic}
\author{R. Grill}
\affiliation{Charles University, Faculty of Mathematics and Physics, Institute
of Physics, Ke Karlovu 5,
 CZ-121~16 Prague 2, Czech Republic}
\author{L. Smr\v{c}ka}
\affiliation{Institute of Physics, Academy of Sciences of the Czech Republic,
Cukrovarnick\'{a} 10, CZ-162 53 Prague 6, Czech Republic}
\author{M. Zv\'{a}ra}
\affiliation{Charles University, Faculty of Mathematics and Physics, Institute of Physics, Ke Karlovu 5,
 CZ-121~16 Prague 2, Czech Republic}

\date{\today}

\begin{abstract}
We present a concept of terahertz oscillations in superlattices
generated under conditions apparently different from standard Bloch
oscillations. The oscillations are induced by crossed magnetic and electric
fields both applied to the superlattice in the in-plane direction. The
frequency of these oscillations is tunable by the applied fields.
\end{abstract}


\maketitle

\section{Introduction}
The possibility of Bloch oscillations (BOs), i.e. the periodic motion of an
electron in a periodic system induced by a uniform electric field, was
mentioned by Zener\cite{Zener} relatively soon after the basic quantum
mechanical theory of the solid state was established.\cite{Bloch} Taking a
simple picture of electrons, which are not subject to tunneling to other bands or
scattering, Zener predicted an oscillatory motion in the real as well as in the reciprocal space
with the frequency $\omega_{BO}=|e|F\Delta/\hbar$, where $\Delta$ and $F$ are
the system spatial period and the electric field, respectively. However, it
took a long time before any experimental evidence of BOs has been found.\cite{Feldman,Waschke}
The key feature for their observation was a
pioneering concept of a superlattice (SL) and minibands suggested by Esaki and
Tsu\cite{Esaki} that allowed to overcome problems of a strong electron
scattering, which made the observation of BOs in bulk semiconductors hardly
feasible.

In this paper, we present a simple idea of electron-in-plane-oscillations
appearing when crossed in-plane magnetic and electric fields are applied to SL.
We start with a general discussion of properties of SL subject to the in-plane
magnetic field the effect of which cannot be described within the
quasi-classical approximation. The findings are illustrated by simple numerical
calculations based on the standard tight-binding (TB) model.\cite{Goncharuk, Lebed}

Subsequently, we study influence of an additionally applied electric field and
conclude that an oscillatory motion of electrons is induced. We present two
models of these oscillations, i) quasi-classical and ii) pure
quantum-mechanical. The predicted oscillations are compared to common BOs.

\section{Superlattice subject to in-plane magnetic field}

Let us consider an infinite SL having its growth axis oriented along the $z$-direction,
which is described by the periodic potential $V(z)=V(z+\Delta)$,
where $\Delta$ is the period of SL. The Hamiltonian of an electron in such a system subject to
the in-plane magnetic field $\mathbf{B}=(0,B_{\|},0)$, with the vector potential
gauge
$\mathbf{A}=(B_{\|}z,0,0)$, reads:
\begin{equation}
\mathcal{H} = \frac{1}{2m} \left(p_x+|e|B_{\|}z \right)^2+\frac{p_y^2 + p_z^2}
{2m} + V(z).
\label{Hamiltonian0}
\end{equation}
To find eigenstates of this Hamiltonian, the following ansatz for the
wave functions is commonly assumed:\cite{Goncharuk,HuPRB92,OrlitaPRB05}
\begin{equation}
\label{psi}
\psi^{}_{k_x,k_y}(x,y,z) = {\rm e}^{ik_xx+ik_yy} \chi^{}_{k_x}(z).
\end{equation}
This way the three-dimensional Hamiltonian \eqref{Hamiltonian0} is reduced to one-dimensional
$H$ which depends on the parameters $k_x$ and $k_y$:
\begin{equation}\label{Hamiltonian1}
H= \frac{\hbar^2}{2m}\left(k_x+\frac{|e|B_{\|}z}{\hbar}\right)^2+\frac{\hbar^2
k_y^2}{2m}- \frac{\hbar^2}{2m}\frac{d^2}{dz^2}+V(z).
\end{equation}
Hence, when the in-plane magnetic field is applied to the SL system,
the motion in the $x$- and $z$-directions become coupled. This coupling can
be interpreted as an effect of the Lorentzian force.

To solve the eigenvalue problem defined by the Hamiltonian~\eqref{Hamiltonian1}, we
can utilize the invariance of this Hamiltonian under transformation
$H(z,k_x)\rightarrow H(z+\Delta,k_x-K_0)$, where $K_0=|e|B_{\|}\Delta/\hbar$.
The whole eigenenergy spectrum then reads:
\begin{equation}
E_n(k_x,k_y) = E_n(k_x) +\frac{\hbar^2k_y^2}{2m},
\label{spec}
\end{equation}
where $E_n(k_x)$ are Landau subbands ($n=1,2,3\ldots$) which are $K_0$--periodic in the momentum $k_x$.

A straightforward solution of the Schr\"{o}dinger equation including the
Hamiltonian~\eqref{Hamiltonian1} leads to the functions $\psi^{(n)}_{k_x,k_y}(x,y,z)$
that do not respect the translation symmetry of the SL along the growth axis. This is caused
by the additional effective potential parabolic in $z$, developed at finite $B_{\|}$
in the Hamiltonian~\eqref{Hamiltonian1}. Consequently, electron momenta $k_x$ have to be
taken within the interval $k_x\in(-\infty,+\infty)$ to obtain full set of eigenfunctions~\eqref{psi}.
Nevertheless, all eigenstates in the $n$-th subband having momenta $k_x+lK_0$ ($l\in\mathbb{Z}$)
are degenerate due to the periodicity of $E_n(k_x)$ and therefore, the
translation periodicity can be restored, when an appropriate linear combination
of these states is taken:
\begin{equation}\label{combination}
\Phi^{(n)}_{k_x,k_y,q}(x,y,z)\propto\sum_{l\in\mathbb{Z}} e^{iql\Delta}\psi^{(n)}_{k_x+lK_0,k_y}(x,y,z).
\end{equation}
The new continuous quantum number $q\in(-\pi/\Delta,\pi/\Delta)$ thus replaces the
discrete index $l$. Owing to the restored translation periodicity, the eigenstates~\eqref{combination}
fulfill the condition:
\begin{equation}\label{condition}
\Phi^{(n)}_{k_x,k_y,q}(x,y,z-\Delta)=e^{i(K_0x+q\Delta)}\Phi^{(n)}_{k_x,k_y,q}(x,y,z)
\end{equation}
and the considered interval of momenta $k_x$ can now be reduced to the Brillouin zone
$k_x\in(-K_0/2,K_0/2)$. Both reduced and extended schemes are fully equivalent.

Hence, the Hamiltonian~\eqref{Hamiltonian1}
describes a system with periodic dispersions $E_n(k_x)$, whose period $K_0$ is
tunable by the applied magnetic field $B_{\|}$. This is in contrast to a fixed
period $2\pi/\Delta$ of the electron dispersion in the growth direction of SL at zero $B_{\|}$.
The obtained result can be interpreted also as a formation of a 2D lattice in
the $x-z$ plane induced in SL by a finite magnetic field in the $y$-direction.
This lattice has spatial periods in $x$- and $z$-directions $2\pi/K_0$ and
$\Delta$, respectively. The magnetic flux trough the unit cell of the lattice
is simply $2\pi B_{\|} \Delta/K_0=h/|e|$, i.e. one magnetic flux quantum.

\begin{figure}
\begin{center}\leavevmode
\scalebox{0.78}{\includegraphics*[104pt,438pt][400pt,703pt]{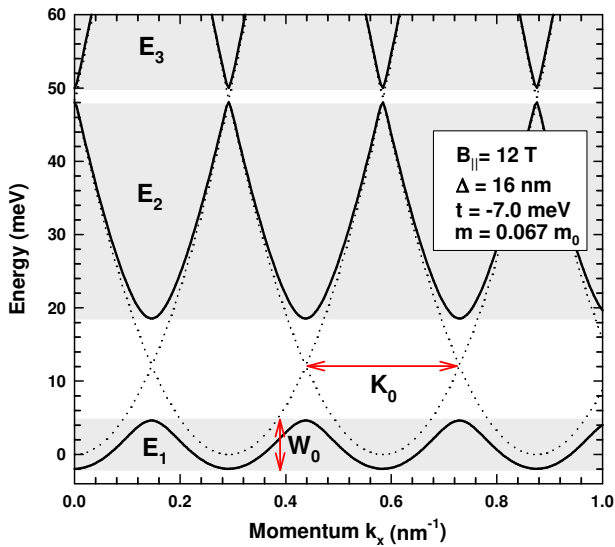}}
\caption{ Electron dispersion curves $E_n(k_x)$ calculated in TB approximation
for three lowest lying subbands $n=1,2,3$. The parameters used in the
calculation are given in the inset.} \label{subbands}
\end{center}
\end{figure}

The form of the Hamiltonian \eqref{Hamiltonian1} implies the appearance of the periodic
dispersion even at a negligible small $B_{\|}$ and thus the standard
parabolic dispersion is not attained in the limit $B_{\|}\rightarrow0+$.
The problem lies in the infinite size of the considered SL.
When a (realistic) superlattice with a finite number of wells is taken into
account, we get only a partially periodic dispersion $E_n(k_x)$ and the number of minima
in this dispersion corresponds to the number of wells, see an extreme case of a double
quantum well in the Appendix. These minima clearly disappear at low $B_{\|}$ and the dispersion
approaches the expectable parabolic shape.

Before we further utilize the above drawn conclusions, we use a simple TB approximation to get some
numerical results, which can illustrate the studied problem. Within the framework of
the TB model, the Hamiltonian~(\ref{Hamiltonian1}) is
transformed into the matrix form and reads:\cite{Wulf,Vyborny}
\begin{equation}\label{TB}
H_{r,s}=\frac{\hbar^2}{2m}(k_x+rK_0)^2\delta_{r,s}+t\delta_{r,s\pm1} \quad (r,s\in\mathbb{Z}) ,
\end{equation}
where the coefficient $t$ $(t<0)$ characterizes the tunneling between adjacent
quantum wells (QWs) and where the motion in the $y$-direction is not included,
since it is not affected by $B_{\|}$. Note that this TB Hamiltonian conserves
the periodicity of the original Hamiltonian \eqref{Hamiltonian1} in momentum
$k_x$. The eigenvalue problem given by the tridiagonal Hamiltonian~\eqref{TB}
can be easily solved numerically.

The calculated dispersions for three lowest lying Landau subbands $E_n(k_x)$
have been plotted into Fig.~\ref{subbands}. These results demonstrate both the expected periodicity of
subbands in $k_x$ and the shape of the dispersion curves, which cannot be predicted only
from the symmetry of the Hamiltonian~(\ref{Hamiltonian1}). We see that just
one minimum per interval $(k_x,k_x+K_0)$ appears. The dotted lines in Fig.~\ref{subbands}
show the limit of very weakly coupled wells, i.e. $t\rightarrow0$. The
dispersion curves at $t=0$ are purely parabolic and corresponds to dispersions of electrons in
isolated QWs. Hence, the widths of individual Landau subbands $E_n(k_x)$ and the energy
gaps between them are given by the strength of $t$ and can be also tuned
by the applied magnetic field $B_{\|}$.

Henceforth, we will take account of the lowest lying subbands $E_1(k_x)\equiv E(k_x)$ only.
Such approximation is meaningful in strongly coupled SLs, i.e. for high values of $|t|$, when
the separation of this lowest subband from the higher ones is significant. The energy
width $W_0$ of this subband can be simply estimated at high magnetic fields. We just take
$W_0\approx\hbar^2(K_0/2)^2/(2m)=e^2B^2_{\|}\Delta^2/8m$. Obviously, this rough approximation
fails if $W_0\approx |t|$.

\section{Semi-classical model of oscillations}

\begin{figure}
\begin{center}\leavevmode
\scalebox{0.4}[0.4]{\includegraphics*[125pt,274pt][500pt,615pt]{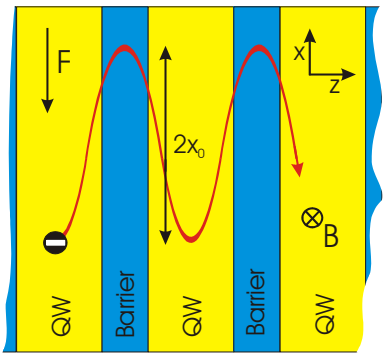}}
\caption{ A schematic picture of SL showing an expected trajectory of the
electron under depicted conditions.} \label{tera2}
\end{center}
\end{figure}

Having the periodic band structure $E(k_x)$ at a given fixed magnetic field
$B_{\|}$, we use a semi-classical consideration to describe the electron motion
if an additional constant electric field $F_x$ is applied in $x$-direction. As
the influence of $B_{\|}$ has already been included in the discussed energy
spectrum, the semiclassical equation of motion takes a simple form $\hbar \dot{k}_x=-|e|F_x$
and thus $k_x$ changes linearly in time. Hence, the electron
velocity $v_x=\hbar^{-1}dE(k_x)/dk_x$ becomes periodic in time and a specific
oscillatory motion is generated. This motion is schematically shown
in~Fig.~\ref{tera2} and can be decomposed into a steady shift in the SL growth
direction and the oscillations in the $x$-direction. The drift motion in the growth-axis
direction becomes apparent especially in the extended scheme of $E(k_x)$. The used equation of motion
gives us also possibility to calculate the corresponding oscillatory frequency
$\omega_{B_{\|}}=2\pi F_x/(B_{\|}\Delta$). Hence, $\omega_{B_{\|}}$ is tunable
not only by the electric field, as in the case of BOs but by $B_{\|}$ as
well. Both frequencies $\omega_{BO}$ and $\omega_{B_{\|}}$ are functions of
$\Delta$ -- but whereas the first frequency is linear in $\Delta$, the latter
one has the reciprocal dependence. The oscillatory frequency $\omega_{B_{\|}}$
can be rewritten into $\omega_{B_{\|}}=2\pi v_d/\Delta$, where $v_d=F_x/B_{\|}$
is the drift velocity introduced in 3D for the electron  motion perpendicular
to the crossed electric and magnetic fields.\cite{Davies} The ratio
$\Delta/v_d$ is then obviously the time needed by an electron to tunnel into the adjacent QW.

The semiclassical model allows us to determine the spatial amplitude of
expected oscillations $x_0$ defined in Fig.~\ref{tera2}. When we make use of
the facts that the electron position is the time integral of the electron
velocity and the velocity $v_x$ is derivative of the dispersion curve, we
obtain a simple relation $x_0=W_0/(2|e|F_x)$. As the subband width $W_0$ varies
with $B_{\|}$, the amplitude $x_0$ is tunable by the magnetic field as well.

\section{Quantum-mechanical model of oscillations}

The electron motion in a system with a periodic dispersion $E(k_x)=E(k_x+K_0)$
can be treated in a pure quantum-mechanical way as reviewed e.g. by Hartmann
\emph{et al.} in Ref.~\onlinecite{Hartmann}. The corresponding Hamiltonian is there
conveniently written in the momentum representation:
\begin{equation}\label{momentum}
H(k_x)=E(k_x)+i |e|F_x\frac{d}{dk_x}
\end{equation}
and thus, taking account of the periodicity in $k_x$, the eigenenergies can be
easily calculated:
\begin{equation}\label{eigen}
E_n=\frac{1}{K_0}\int_0^{K_0}E(k_x)dk_x+n\hbar\omega_{B_{\|}}.
\end{equation}
Because the first term of this eigenenergy is constant at given $B_{\|}$, we
receive an analog of the common Wannier-Stark ladder ($n\in \mathbb{Z}$)
discussed in the framework of Bloch oscillations.\cite{Hartmann}

\section{Remarks on a possible realization}

From a practical point of view, the experiments proving emission of the
predicted THz radiation can be the same as in the case of standard
BOs. Both coherent and incoherent radiations can be obtained. The coherent THz radiation induced by BOs is achieved
e.g. when free electrons are generated in an undoped SL by a femtosecond optical pulse
ensuring the same phase of all electrons.\cite{Feldman,Waschke,Martini}
The photon laser energy is tuned to an
appropriate electron-hole transition in SL. The same technique can be used in
our case as well.

The function of the THz generator could be disrupted, if electrons
initially localized in the lowest subband tunnel under the effect
of $F_x$ to the higher subbands. In such a case, the one-subband model
utilized in both semiclassical and quantum-mechanical treatments of
oscillations would not be applicable. We illustrate this obstacle on a
simple model of a double quantum well in Appendix. This model
offers the simplest possibility to check the intersubband tunneling induced by
the electric field. It cannot serve as a definite evidence that the tunneling
to higher subbands is negligible in superlattices, nevertheless, it
illustrates that electrons do not noticeably tunnel to the higher (antibonding) subband
in DQW at $B_{\|}$ under conditions typical for the THz oscillations
predicted in SLs. Apparently, further investigations
in this direction are necessary.

For a possible realization, we should also check the sample design and
experimental conditions to observe the predicted oscillations in the terahertz
region. Assuming $\omega_{B_{\|}}\approx\omega_{BO}$ and the same electric
field in both cases, we obtain the corresponding magnetic field $B_{\|}\approx
h/\Delta^2|e|\cong16$~T for $\Delta=16$~nm. Hence, at $B_{\|}<16$~T the
oscillatory frequency is even higher than for BOs, since
$\omega_{B_{\|}}\propto B_{\|}^{-1}$. Moreover, having two free parameters
$F_x$ and $B_{\|}$ we can independently optimize $\omega_{B_{\|}}$ and $x_0$ to
achieve the maximal emitted power. This is impossible for standard BOs,
since $\omega_{BO}$ and the corresponding spatial amplitude are governed by the
applied electric field only.

The important point in the observation of BOs is the achievement of an
oscillation period significantly lower than is the scattering time due to
phonons or plasmons. We predict our oscillation for SL systems, where common
BOs are observed. Therefore, the same or very similar damping rates as observed
in BO experiments could be expected in our case as well. Hence, the published
experimental evidence of BOs,\cite{Feldman,Waschke,Martini} suggests
that the predicted $B_{\|}$-controlled oscillations ought to be experimentally observable.

It is interesting to investigate also the direction characteristics of the
expected THz radiation. Since the radiation is generated by the electron
oscillatory motion in the $x$-direction, the radiation should be emitted mainly
in the plane perpendicular to the $x$-axis, i.e. in the plane perpendicular to
the oscillating dipoles. The predicted device can thus be both edge- or
surface-emitting.

An important advantage of presented model in comparison with
standard BOs is a fast (in-plane) drain of electrons from the
structure after they reach the edge of SL. This fact can be
utilized in a significant enhancement of the repetition
frequency of the generation of the coherent THz radiation.

\section{Conclusions}

We have investigated behavior of electrons in a superlattice when crossed
magnetic and electric fields are applied, both in the in-plane direction. We
predict a novel terahertz oscillations in superlattices that are different from
Bloch oscillations that appear when the electric field is applied in the
growth direction of the superlattice. We have also found a simple expression
for the frequency of such oscillations. The suggested realistic design of the
structures allows preparation of terahertz emitters controlled by the in-plane
magnetic field.

\section{Appendix}

\begin{figure}[t]
\begin{center}
\scalebox{0.75}{\includegraphics*[151pt,322pt][428pt,523pt]{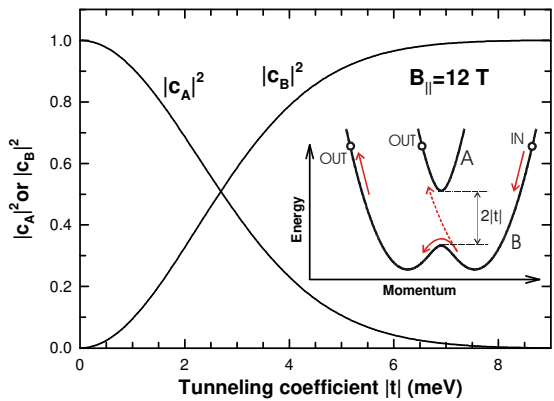}}
\caption{\label{THzInter} The occupancy $|c_B|^2$ ($|c_A|^2$) of
the bonding (antibonding) subband in DQW after the reflection of
an electron on a barrier formed by the lateral in-plane electric
field. The inset schematically illustrates the coming (IN) and reflected (OUT) states
in the electron subband structure corresponding to the region $x<0$.}
\end{center}
\end{figure}

Let us assume a double quantum well (DQW) oriented as used
in the paper subject to the in-plane magnetic field and described within the
TB approximation by the interwell distance $\Delta$ and the tunnelling coefficient
$t$. The thorough theoretical analysis of such a DQW system can be found
elsewhere.\cite{HuPRB92,LyoPRB94} The DQW represents an extreme case of a
superlattice taken up to now into account. A lateral electric field applied
in the $x$-direction at $x>0$ forms the potential profile:
\begin{eqnarray}
\label{VCdx}
\phi(x)=\left\{ \begin{array}{ll}
0 & \mbox{if $x<0$}\\
|e|F_xx & \mbox{if $x>0$} \end{array}\right..
\end{eqnarray}
The electron described by the bonding subband wave function
$\Psi_B(k_x,x)$ with the momentum $k_x>0$ and energy $E_B(k_x)$
enters into the system at $x<0$. It is reflected by the potential
barrier $\phi(x)$ and leaves the system again at $x<0$. The
reflected electrons may be found both in bonding states
$\Psi_B(-k_x,x)$ and in antibonding states $\Psi_A(-k'_x,x)$ with
$k'_x>0$ being a momentum of the corresponding antibonding state.
Due to the elastic reflection the energy of antibonding state
$E_A(-k'_x)$ equals to bonding ones $E_A(-k'_x)=E_B(\pm k_x)$.

The total wave function $\Psi(x)$ of the electron at $x<0$ thus
reads
\begin{eqnarray}
\label{Psi}
\Psi=\frac{\Psi_B(k_x)+c_B\Psi_B(-k_x)+c_A\Psi_A(-k'_x)}{\sqrt{1+c_Bc_B^*+c_Ac_A^*}},
\end{eqnarray}
where the complex amplitudes of reflected waves $c_B$ and $c_A$
determine the occupancy of respective bonding and antibonding
subbands. The wave function $\Psi$ at $x>0$ is calculated solving
the Schr{\"o}dinger equation numerically and respective $c_B$ and
$c_A$ are established to accomplish the damping
$\lim_{x\rightarrow+\infty}\Psi(x)=0$. Results of
the calculation for different tunnelling coefficients $t$ are shown in Fig.
\ref{THzInter}, where model parameters $m=0.067m_0$, $\Delta=16$~nm,
$B_{\|}=12$~T and
$F_x=1920$~V/cm corresponding to the oscillator
frequency of 1~THz have been used. The energy of incoming electrons $E_B(k_x)$ well above
the minimum energy of the antibonding subband $E_A(0)$ was used to
enable the intersubband tunneling. We have ascertained that
the course of $|c_B|^2$ ($|c_A|^2$) is practically insensitive
to $E_B(k_x)$. We find out that increased subband
splitting strongly damps the intersubband tunnelling and
$|c_A|^2<0.01$ is obtained at $|t|>7.5$ meV for chosen parameters. Analogous results are
obtained also for other sets of parameters producing oscillations
in THz branch.

\begin{acknowledgments}
This work is a part of the research plan MSM 0021620834 that is financed by the
Ministry of Education of the Czech Republic. M. O. acknowledges the support
from Grant Agency of Charles University under contract No.~281/2004 and L. S.
from Grant Agency of ASCR under contract No.~IAA1010408.
\end{acknowledgments}


\begin{thebibliography}{10}
\bibitem{Zener} C. Zener, Proc. R. Soc. London A 145, (1934) 523.
\bibitem{Bloch} F. Bloch, Z. Phys. 52, (1928) 555.
\bibitem{Feldman} J. Feldmann, K. Leo, J. Shah, D. A. B. Miller, J. E. Cunningham, T. Meier, G.
von Plessen, A. Schulze, P. Thomas, and S. Schmitt–-Rink, Phys. Rev. B 46,
(1992) 7252.
\bibitem{Waschke} C. Waschke, H. G. Roskos, R. Schwedler, K. Leo, H.
Kurz, and K. K\"{o}hler, Phys. Rev. Lett. 70, (1993) 3319.
\bibitem{Esaki} L. Esaki and R. Tsu, IBM J. Res. Dev. 14, (1970) 61.
\bibitem{Goncharuk} N. A. Goncharuk, L. Smr\v{c}ka, J. Ku\v{c}era, and K.
V\'{y}born\'{y}, Phys. Rev. B 71, (2005) 195318.
\bibitem{Lebed} A. G. Lebed, Phys. Rev. Lett. 95, (2005) 247003.
\bibitem{HuPRB92} J. Hu and A. H. MacDonald, Phys. Rev. B 46, (1992) 12554.
\bibitem{OrlitaPRB05} M. Orlita, R. Grill, P. Hl\'{\i}dek, M. Zv\'{a}ra, G. H.
{D\"{o}hler}, S. Malzer, and M. Byszewski, Phys. Rev. B 72, (2005) 165314.
\bibitem{Wulf}  U. Wulf, J. Ku\v{c}era, and A. H. MacDonald, Phys. Rev. B 47, (1993) 1675.
\bibitem{Vyborny} K. V\'{y}born\'{y}, L. Smr\v{c}ka, and R. A. Deutschmann, Phys. Rev. B 66, (2002) 205318.
\bibitem{Davies} J.~H.~Davies,~The~Physics~of~Low-Dimensional Semiconductors: An Introduction, Cambridge
University Press, (1997) p. 229.
\bibitem{Hartmann} T. Hartmann, F. Keck, H. J. Korsch, and S. Mossmann, New J. Phys. 6, (2004) 2.
\bibitem{Martini} R. Martini, G. Klose, H. G. Roskos, H. Kurz, H. T. Grahn, and R.
Hey, Phys. Rev. B 54, (1996) R14325.
\bibitem{LyoPRB94} S. K. Lyo, Phys. Rev. B 50, (1994) 4965.
\end{thebibliography}
\end{document}